\documentclass[12pt]{article}
\input epsf.sty

\begin{document}

\def\CA{{\cal A}}
\def\CB{{\cal B}}
\def\CC{{\cal C}}
\def\CD{{\cal D}}
\def\CE{{\cal E}}
\def\CF{{\cal F}}
\def\CG{{\cal G}}
\def\CH{{\cal H}}
\def\CI{{\cal I}}
\def\CJ{{\cal J}}
\def\CK{{\cal K}}
\def\CL{{\cal L}}
\def\CM{{\cal M}}
\def\CN{{\cal N}}
\def\CO{{\cal O}}
\def\CP{{\cal P}}
\def\CQ{{\cal Q}}
\def\CR{{\cal R}}
\def\CS{{\cal S}}
\def\CT{{\cal T}}
\def\CU{{\cal U}}
\def\CV{{\cal V}}
\def\CW{{\cal W}}
\def\CX{{\cal X}}
\def\CY{{\cal Y}}
\def\CZ{{\cal Z}}

\newcommand{\todo}[1]{{\em \small {#1}}\marginpar{$\Longleftarrow$}}
\newcommand{\labell}[1]{\label{#1}\qquad_{#1}} 
\newcommand{\bbibitem}[1]{\bibitem{#1}\marginpar{#1}}
\newcommand{\llabel}[1]{\label{#1}\marginpar{#1}}

\newcommand{\sphere}[0]{{\rm S}^3}
\newcommand{\su}[0]{{\rm SU(2)}}
\newcommand{\so}[0]{{\rm SO(4)}}
\newcommand{\bK}[0]{{\bf K}}
\newcommand{\bL}[0]{{\bf L}}
\newcommand{\bR}[0]{{\bf R}}
\newcommand{\tK}[0]{\tilde{K}}
\newcommand{\tL}[0]{\bar{L}}
\newcommand{\tR}[0]{\tilde{R}}

\newcommand{\btzm}[0]{BTZ$_{\rm M}$}
\newcommand{\ads}[1]{{\rm AdS}_{#1}}
\newcommand{\ds}[1]{{\rm dS}_{#1}}
\newcommand{\eds}[1]{{\rm EdS}_{#1}}
\newcommand{\sph}[1]{{\rm S}^{#1}}
\newcommand{\gn}[0]{G_N}
\newcommand{\SL}[0]{{\rm SL}(2,R)}
\newcommand{\cosm}[0]{R}
\newcommand{\hdim}[0]{\bar{h}}
\newcommand{\bw}[0]{\bar{w}}
\newcommand{\bz}[0]{\bar{z}}
\newcommand{\be}{\begin{equation}}
\newcommand{\ee}{\end{equation}}
\newcommand{\bea}{\begin{eqnarray}}
\newcommand{\eea}{\end{eqnarray}}
\newcommand{\pat}{\partial}
\newcommand{\lp}{\lambda_+}
\newcommand{\bx}{ {\bf x}}
\newcommand{\bk}{{\bf k}}
\newcommand{\bb}{{\bf b}}
\newcommand{\BB}{{\bf B}}
\newcommand{\tp}{\tilde{\phi}}
\hyphenation{Min-kow-ski}

\def\apr{\alpha'}
\def\str{{str}}
\def\lstr{\ell_\str}
\def\gstr{g_\str}
\def\Mstr{M_\str}
\def\lpl{\ell_{pl}}
\def\Mpl{M_{pl}}
\def\varep{\varepsilon}
\def\del{\nabla}
\def\grad{\nabla}
\def\tr{\hbox{tr}}
\def\perp{\bot}
\def\half{\frac{1}{2}}
\def\p{\partial}
\def\perp{\bot}
\def\eps{\epsilon}
\newcommand{\Tr}{\mathop{\rm Tr}}


\def\NN{{\cal N}}
\def\nfour{{\cal N}=4}
\def\ntwo{{\cal N}=2}
\def\none{{\cal N}=1}
\def\nonestar{{\cal N}=1$^*$}
\def\tr{{\rm tr\ }}
\def\RR{{\cal R}}
\def\PP{{\cal P}}
\def\ZZ{{\cal Z}}

\newcommand{\bel}[1]{\be\label{#1}}
\def\al{\alpha}
\def\bt{\beta}
\def\mn{{\mu\nu}}
\newcommand{\rep}[1]{{\bf #1}}
\newcommand{\vev}[1]{\langle#1\rangle}
\def\bra{\langle}
\def\ket{\rangle}
\def\eref{(?FIX?)}

\renewcommand{\thepage}{\arabic{page}}
\setcounter{page}{1}

\def\Tbar{\bar{T}}
\def\pbar{\bar{\partial}}
\def\psibar{\bar{\psi}}

\rightline{hep-th/0202187}
\rightline{UPR-981-T,  HIP-2002-07/TH}
\vskip 1cm
\centerline{\Large {\bf A Space-Time Orbifold:}}
\vskip 0.3cm
\centerline{\Large {\bf  A Toy Model for a Cosmological
Singularity}}
\vskip 0.5cm
\renewcommand{\thefootnote}{\fnsymbol{footnote}}
\centerline{{\bf Vijay
Balasubramanian,$^{1}$\footnote{vijay@endive.hep.upenn.edu}
S. F. Hassan,$^{2}$\footnote{fawad.hassan@helsinki.fi}}}
\centerline{{\bf Esko Keski-Vakkuri,$^{2}$\footnote{keskivak@rock.helsinki.fi} and
Asad Naqvi$^{1}$\footnote{naqvi@rutabaga.hep.upenn.edu}
}}
\vskip .5cm
\centerline{\it {}$^{1}$David Rittenhouse Laboratories, University of
Pennsylvania}
\centerline{\it Philadelphia, PA 19104, U.S.A.}
\vskip .5cm
\centerline{\it ${}^{2}$Helsinki Institute of Physics,
} \centerline{\it P.O.Box 64, FIN-00014 University of Helsinki, Finland}

\setcounter{footnote}{0}
\renewcommand{\thefootnote}{\arabic{footnote}}

\begin{abstract}
We explore bosonic strings and Type II superstrings in the simplest
time dependent backgrounds, namely orbifolds of Minkowski space by
time reversal and some spatial reflections.  We show that there are no
negative norm physical excitations.  However, the contributions of
negative norm virtual states to quantum loops do not cancel, showing
that a ghost-free gauge cannot be chosen.  The spectrum includes a
twisted sector, with strings confined to a ``conical'' singularity
which is localized in time.  Since these localized strings are not
visible to asymptotic observers, interesting issues arise regarding
unitarity of the S-matrix for scattering of propagating states.  The
partition function of our model is modular invariant, and for the
superstring, the zero momentum dilaton tadpole vanishes.  Many of the
issues we study will be generic to time-dependent cosmological
backgrounds with singularities localized in time, and we derive some
general lessons about quantizing strings on such spaces.
\end{abstract}

\section{Introduction}

Time-dependent space-times are difficult to study, both classically
and quantum mechanically. For example, non-static solutions are harder
to find in General Relativity, while the notion of a particle is
difficult to define clearly in field theory on time-dependent
backgrounds. Quantum mechanical strings propagating on time-dependent
spaces can develop many subtle problems including difficulties with
unitarity and ghosts in the physical spectrum. Nevertheless, the apparent
observation of a cosmological constant from supernovae
measurements~\cite{supernova}, and an attendant expansion of the
universe, requires us to understand clearly how time dependence of
cosmological backgrounds is incorporated into string theory. In
related theoretical developments, recent work has explored the physics
of de Sitter space~\cite{desitter}, as well as new pictures of the
early universe in which a collision of branes forms the observable
cosmic structures~\cite{ekpyrosis}. In the latter models, and in the
pre-big bang scenarios~\cite{Veneziano:2000pz}, a stringy resolution
of an initial singularity is proposed to permit an extension of
space-time to an era before the big bang. In view of all this it is
worthwhile to investigate perturbative string theory in singular
cosmological backgrounds.

Perturbative string theory is most easily studied in flat,
translationally invariant space. The simplest non-homogeneous spaces
in which it is well-defined are orbifolds of flat space in which some
Euclidean directions are quotiented by a discrete subgroup of the
isometry group~\cite{orbifolds}. When the action of the discrete group
has fixed points, the orbifold has conical singularities, as well as
new light states (the so-called twisted sectors) which are confined to
these defects. Condensing twisted sector states can resolve the
conical singularities in many cases such as the classic example
$R^{4}/Z_{2}$ where four Euclidean directions are identified under
reflections.

Can we find consistent backgrounds in string theory by identifying
points in space-time rather than just in space?  One simple example is
the BTZ black hole of three dimensional gravity which is obtained by
quotienting $\ads{3}$ by a boost~\cite{Banados:1992wn}.\footnote{The
consistency of string theory on $\ads{3}$ itself is nontrivial, {\em
e.g.} for the no-ghost theorem and modular invariance see~\cite{ads}. 
For work on string theory in BTZ black holes, see {\em e.g.}
\cite{btz}.} Such orbifolds bear a relation to the kinds of
identifications discussed in the context of resolving singularities
separating contracting and expanding phases of some cosmological
models~\cite{ekpyrosis}.  Likewise, some coset WZW models are
consistent time-dependent string backgrounds \cite{Nappi:1992kv}. 
Also, string theory on orbifolds with time identified under $t
\rightarrow t + 1$ (i.e., circular time) has been studied
in~\cite{moore} and the resulting time-like T-duality has
been studied in \cite{hull}. Space-time singularities in
string theory were studied in \cite{Horowitz:ap}. 
 In this paper, we will seek simple models of
time-dependent spaces and of cosmological singularities by
constructing space-time orbifolds in which we identify space-time
under both time reversal and reflections in some directions. 
Generally speaking, string theories defined on such spaces are
threatened by a number of pathologies including potential ghosts in
the physical spectrum and problems with unitarity. In fact, all known
proofs of the no-ghost theorem explicitly require time-independent
backgrounds~\cite{Asano:2000fp}. Also, supersymmetry is generally
broken and so there may be a danger of tadpoles at one loop and
instabilities like tachyons could occur. Part of our goal is to
explore the many subtleties that beset such constructions in string
theory.

We study bosonic and Type II superstrings on $R^{1,d}/Z_{2}$, in which
we have identified space-time by time reversal and reflections. When
$d=0$, only time is identified and the space has an initial
singularity at $t=0$. When $d\geq 1$ the background geometry is a
space-time cone with a ``conical'' singularity at $t = x_{1}=\cdots=
x_{d}=0$. String theory on such spaces is defined by projecting onto
the sector of the Hilbert space that is invariant under these discrete
transformations, and including possible twisted sectors localized at
the orbifold fixed point at $t = x_{1}=\cdots= x_{d}=0$, and which
therefore do not propagate. After this projection, quantum mechanics
is consistent with closed time-like loops in the geometry. 
We find that the physical states are
ghost-free when $d+1 \geq 9$ for the bosonic string and that there is
no restriction on $d$ for the superstrings. In Type II superstrings,
when $d+1=4$, there is a ``massless'' twisted sector in which physical
states satisfy the on-shell condition $|\vec{p}|^{2} =
0$.\footnote{Actually, this implies that $\vec{p}=0$ since the twisted
sector states are localized in time and so only carry momenta in the
un-orbifolded Euclidean directions.} It is possible that condensing
these states would resolve the conical singularity, and so we focus on
the $d+1=4$ case.

We compute the partition function when $d+1=4$ and find that it is
zero.  Likewise the one loop zero-momentum tadpoles vanish suggesting
that we have a consistent string background at this order in string
perturbation theory.    However, negative norm states (although not
present in the on-shell physical spectrum) make a contribution to the
partition function -- their virtual effects do
not cancel between the matter and ghost sectors as they do in the
standard $R^{4}/Z_{2}$ orbifold.  This shows that it is not possible 
to choose a ghost-free gauge in which all computations are carried out 
in terms of positive norm states.   We expect that this will be 
generally true for string theory in time-dependent backgrounds.\footnote{
This is reminiscent of mixing between the matter and
ghost sector CFTs discussed in \cite{witten}.}

We conclude the paper by discussing several novel subtleties
introduced by the localization in time of a sector of physical states,
and by summarizing lessons learned from our work about time-dependent
backgrounds and cosmological singularities in string theory.

\section{Space-time orbifolds}

We study space-time orbifolds constructed by identifying Minkowski
space under time reversal and reflections in some spatial directions. 
As we will see below, the resulting geometry can be interpreted as a
space-time cone.  After the identifications the covering space has
some closed time-like loops.  However, because the orbifold
prescription projects onto states in Hilbert space that are symmetric
under the identifications, quantum mechanical evolution remains
consistent.

\begin{figure}
  \begin{center}
 \epsfysize=2in
   \mbox{\epsfbox{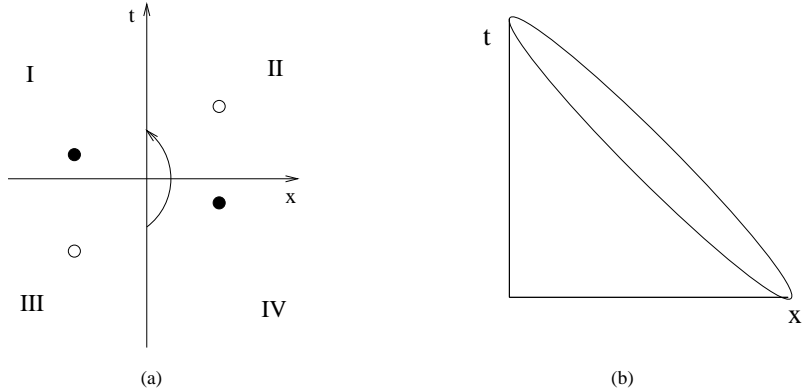}}
\end{center}
\caption{A space-time cone.}
\label{cone}
\end{figure}

\subsection{Conical space-time geometries}
Consider identifying time ($X^0$) and $d$ spatial directions under
reflections:
\begin{equation}
X^a \rightarrow -X^a \qquad  (a=0,\cdots, d)\,,
\end{equation}
leaving all other directions unaffected.  Fig.~\ref{cone} shows the
resulting space-time cone when $d=1$. Points in opposite quadrants of
the $X-T$ plane are identified as in Fig.~\ref{cone}.  Therefore the
quadrants II and IV (or I and II) may be taken as ``fundamental''
regions with independent physics.  Identifying these regions along the
T (or X) axis produces the cone in Fig.~\ref{cone}b with a singular point
at $T=X=0$.

\begin{figure}
  \begin{center}
 \epsfysize=2in
   \mbox{\epsfbox{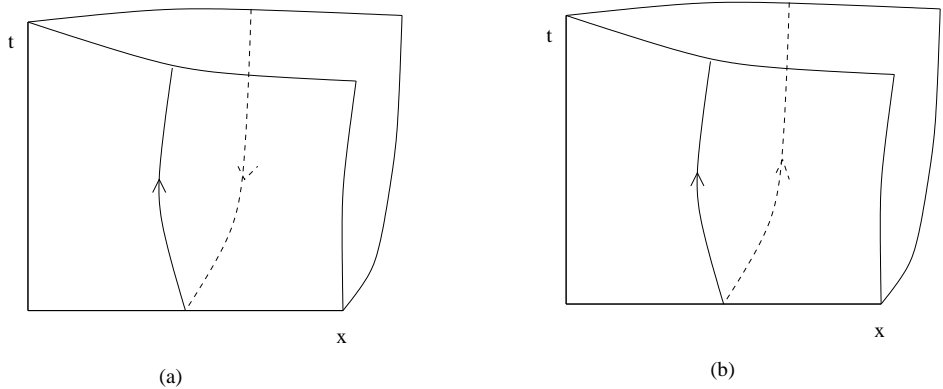}}
\end{center}
\caption{Time evolution on the cone.}
\label{pocket}
\end{figure}

The proper distance on the covering space between a point $(T,X)$ and
its image $(-T,-X)$ is $ds^2=4(X^2-T^2)$. This is time-like in the
region inside the light cone emanating from the point $T=X=0$ on the
covering space. As a result there are closed time-like curves in this
geometry, such as the one in Fig.~\ref{cone}a. In the orbifold
construction which we will describe below, such loops do not pose a
fundamental problem since we are instructed to project to states in
the Hilbert space that are invariant under the space-time
identifications, i.e., we project onto quantum mechanical
wave-functions that satisfy $\psi(x,t) =\psi(-x,-t)$. A picture of
time evolution on the cone is provided in Fig.~\ref{pocket}a where we
have folded regions II and IV along the $X$-axis and identified the
negative and positive directions along the time axis, to make a cone.
It is natural then to describe the evolution of states on the cone
with respect to the time direction inherited from the positive time
direction in quadrants II and IV of the parent manifold. The line
$x=0$ appears to have time ``running both ways'', but this is simply
because we have projected onto states that are time reversal invariant
on the $X=0$ axis.

Constructing the cone by gluing the X axis of quadrants I and II
yields a similar picture with two ``sheets'' glued together on the T
and X axis.  At first sight the time inherited from the covering space
gives evolution moving ``up'' on both sheets in Fig~\ref{pocket}b,
with the boundary condition that the wave-functions on both sheets
approach the same value on a big-bang-like surface at $T=0$.  However,
on the X axis of the covering space the orbifold identifications also
imply that $\partial\psi(x,t)/\partial t|_{t=0} = - \partial\psi(-x,
t)/\partial t|_{t=0}$.  Therefore, on the cone, with time evolving
``up'' on both sheets, although wave-functions on both sheets agree on
the initial surface, their time derivatives are opposites of each
other.  Therefore it seems more natural once again to describe the
evolution of states with respect to a continuous time as in
Fig.~\ref{pocket}a.

\subsection{Euclidean world-sheets and Lorentzian backgrounds}
As we have discussed, we will construct string theory on our
space-time orbifold by projecting onto states of strings in Minkowski
space that are invariant under the discrete identifications. In
Lorentzian space-times the signature of the string worldsheet must be
$(-1,1)$ in order for classical string propagation to exist. (The $2d$
equations of motion are solved by equating the worldsheet metric with
the metric induced from space-time.) Nevertheless, the standard
techniques of string theory involve analytically continuing the
worldsheet to Euclidean signature in order to exploit the techniques
of two-dimensional conformal field theory and complex geometry. In
static backgrounds we might imagine continuing the space-time to
Euclidean signature at the same time, but this is not possible in
time-dependent backgrounds such as ours. Our analysis in this paper
is done with a Lorentzian signature world sheet except our discussion of 
modular invariance where we  formally continue the world sheet to
Euclidean signature. The resulting path integral
(with a Euclidean world sheet and Lorentzian target space) is not
strictly speaking well-defined because the action is not bounded from
below.  Nevertheless, it appears to be finite in our case, and 
we use it formally to discuss modular invariance. Subtleties in
defining the Polyakov path integral in Lorentzian signature
have been discussed by Mathur in~\cite{Mathur}.

\section{Bosonic string theory on the Lorentzian orbifold }

Before studying superstrings on space-time orbifolds we examine the
$26$-dimensional bosonic string propagating on $R^{1,d}/Z_2$. This
already contains the distinctive features of the Lorentzian orbifold.
In particular, we show that it is possible to obtain a ghost-free
physical spectrum and a modular invariant partition function, but
that virtual negative norm states make un-cancelled contributions
to quantum loops. This is a reflection of the time dependence of 
the string theory background. 

Consider flat $26$-dimensional Minkowski space with points
identified under the $Z_2$ action,
\begin{equation}
X^a \rightarrow -X^a \quad (a=0\cdots d)\,;\quad\quad
X^i \rightarrow X^i \quad (i=d+1 \cdots 25).
\label{orbifold}
\end{equation}
This action has a fixed $(25-d)$-dimensional hyper-plane, given by $X^a=0$.
To get consistent string propagation on this space-time, we
project the conventional bosonic string Hilbert space onto its $Z_2$
invariant subspace. This gives the untwisted sector of the orbifold
theory. In addition, there is a twisted sector corresponding to
strings that are closed only under the identifications made by the
orbifold group. Again, we project out twisted sector states that are
not invariant under the orbifold action. The twisted strings are
trapped around the tip of the cone in Fig.~\ref{cone}b, which is a
$(25-d)$-dimensional hyper-plane localized at an instant in time. The
untwisted strings can propagate in the bulk.

The orbifold above has the novel feature that it includes a reflection
in the time direction, destroying the global time-like isometry of
flat space-time. This means that we cannot perform quantization by
going to light-cone gauge. The alternative is to use the covariant
BRST formalism. However, in the absence of a light-cone gauge choice,
the absence of negative-norm states in the physical spectrum is no
longer evident, especially in view of the non-applicability of the
known proofs of no-ghost theorem \cite{Asano:2000fp}. In the following,
we will mostly be concerned with this issue. In the covariant
formalism, we work with world-sheet fields $X^\mu\, (\mu=0,\cdots,
25)$ and the reparameterization ghosts $b$ and $c$. In the untwisted
sector $X^\mu(\sigma+2\pi,\tau)=X^\mu(\sigma,\tau)$, and the mode
expansion is\footnote{We will work in $\alpha'=2$ units in this
paper.}
\begin{equation}
X^\mu=x^\mu+p^\mu\tau+i\sum_{n\neq 0}{\alpha_n^\mu\over n}
e^{-in(\tau-\sigma)}+i\sum_{n\neq 0}{\tilde{\alpha}_n^\mu\over n}
e^{-in(\tau+\sigma)}\,.
\label{xunt}
\end{equation}
The (tachyonic) ground state $| p^a,p^i \ket$ carries
momentum in both orbifolded and un-orbifolded directions and the
Hilbert space of states is constructed by acting with creation
operators on the ground state. Half of the states with non-zero $p^a$
are projected out of the spectrum. For example, of the states
$\alpha_{-1}^a \tilde{\alpha}_{-1}^i |p^a,p^i\ket$ and $\alpha_{-1}^a
\tilde{\alpha}_{-1}^i |-p^a,p^i\ket$, only the linear combination
$\alpha_{-1}^a\tilde{\alpha}_{-1}^i(|p^a,p^i\ket-|-p^a,p^i\ket)$ is
retained. When $p^a=0$, only the $Z_2$ invariant combinations of the
oscillators acting on the vacuum are kept. Hence,
$\alpha_{-1}^a\tilde{\alpha}_{-1}^i|0,p^i\ket$ is projected out but
$\alpha_{-1}^a\tilde{\alpha}_{-1}^b|0,p^i\ket$ is retained.

In the twisted sector, the fields $X^a$ satisfy the anti-periodic
boundary condition $X^a(\sigma+2\pi)=-X^a(\sigma)$, with the mode
expansion given by,
\begin{equation}
X^a=i\sum_{r}{\alpha^a_r\over{r}}e^{-ir(\tau-\sigma)}+
i\sum_{r}{\tilde{\alpha}^a_r\over{r}}e^{-ir(\tau+\sigma)}\,,
\label{xtwist}
\end{equation}
where $r$ is half odd-integral. The twisted sector is localized at the
orbifold fixed plane at $X^a=0$. In particular, for any value of the
parameter $\tau$, the twisted string world-sheet does not propagate
too far out in time $X^0$. This is similar to an instanton. The mode
expansion for $X^i$ in directions transverse to the orbifold is the
same as in the untwisted sector. Consequently, the ground state in the
twisted sector carries a momentum $p^i$ only in the transverse
directions. The Hilbert space is built by acting with creation
operators on the ground state and projecting onto the $Z_2$ invariant
subspace.

The ghosts $b$ and $c$ are not affected by the orbifold and 
have the same mode expansions in both sectors: 
\begin{equation}
b(\sigma,\tau)=\sum_n b_n e^{-in(\tau - \sigma)}\,,\qquad
c(\sigma,\tau)=\sum_n c_n e^{-in(\tau - \sigma)}\,;
\end{equation}
and similarly for right-movers $\tilde{b}$ and $\tilde{c}$. 
\subsection{Physical states}
The BRST operator $Q_B$
is given by
\begin{equation}
Q_B=\sum_n(c_nL_{-n}^m+\tilde{c}_n\tilde{L}_{-n}^m)
+\sum_{m,n}{(m-n) \over 2}:(c_mc_nb_{-m-n}+\tilde{c}_{m}\tilde{c}_{n}
\tilde{b}_{-m-n}):+a\, (c_0+\tilde{c}_0),
\label{QB}
\end{equation}
where $L^m$ are the Virasoro generators in the matter sector and $a$
is the zero point energy. Physical states are elements of the BRST
cohomology, {\em i.e.}, they obey $Q_B|\psi\ket=0$ subject to the
equivalence relation $|\psi \ket \sim |\psi \ket + Q_B|\phi \ket$,
where $|\phi \ket$ is an arbitrary state.

It is now easy to see that the physical spectrum does not contain
negative norm states. In the untwisted sector, after the orbifold
projection, the states form a subspace of the Fock space of the
parent theory. Furthermore , the orbifold action
(\ref{orbifold}) commutes with the BRST operator (\ref{QB}). This
means that the space of physical states of the orbifold theory is 
a subspace of the space of physical states of the parent theory, and
hence is free of negative norm states. More explicitly, for $p^a\neq
0$, one can easily establish a correspondence between states in the
parent theory and those of the orbifold theory by appropriately
choosing symmetrized or antisymmetrized momentum wave-functions. 

To see that the twisted sector physical states do not have
negative norms, recall that the BRST condition $Q_B|\psi \ket =0$, along
with $b_0|\psi\ket=0$, implies (see, for example, \cite{Polchinski:rq})
\begin{equation}
(L_0^m+L_0^{gh}-a) | \psi \ket =0\,,
\end{equation}
where $L_0^{gh}$ is the ghost Virasoro generator. In terms of the
twisted sector number operators, we have 
\begin{equation}
L_0^m+L_0^{gh}= \frac{1}{2} p^ip_i+
\sum_{n=1}^\infty n(N_{bn}+N_{cn}+\sum_{i=d+1}^{25} N_{i n})
+\sum_{r=1/2}^\infty \sum_{a=0}^{d} rN_{a r}\,,
\end{equation}
and the twisted sector zero-point energy is $a=\frac{26-(d+1)-2}{24}
-\frac{d+1}{48}={{15-d}\over 16}$. Then for a physical state
$|\psi,p^i\ket$, with momentum $p^i$ in the un-orbifolded
directions, this implies
\begin{equation}
\frac{1}{2} p^ip_i+\sum_{n=1}^\infty n(N_{bn}+N_{cn}+\sum_{i=d+1}^{25}
N_{i n})+\sum_{r=1/2}^\infty\sum_{a=0}^{d} rN_{a r}
={15-d \over 16}.
\end{equation}

Since the left hand side is always positive, $d$ is restricted to
$d\le 15$ in order to allow for any physical states in the twisted
sector. Furthermore, since $(15-d)/16 < 1$, a twisted sector physical
state will not contain $c$, $b$ and $X^i$ excitations. For $1\le d\le
7$, the physical spectrum will always contain a negative norm state
corresponding to $\alpha^0_{-1/2}$. However,
for $d \geq 8$ there are no negative norm states in the twisted sector
physical spectrum which, for $15 \geq d \geq 8$, contains only the
ground state $|0,p^i\ket$. In particular, $p^i=0$ for $d=15$.
\footnote{For $d=15$, the state in the twisted sector is physical only
when $p^i=0$. If this state at $p^i \neq 0$ were BRST exact, it would
be orthogonal to all other physical states. Amplitudes involving such
a state would then have to be proportional to $\delta^{(9-d)}(p^i)$. As
argued in \cite{Polchinski:rq}, since amplitudes in field theory and
string theory never have this kind of a behaviour, such a state with
$p^i=0$ should not be part of the physical spectrum. 
However, this is {\em not} true for the twisted sector state on the
Lorentzian orbifold. This is because the state with non-zero $p^i$ is
not BRST exact since it is not even BRST closed ($Q_B|p^i\neq0\ket
\neq 0$), as it does not satisfy the Virasoro constraint. So the above
argument does not apply and the zero-momentum physical twisted state
should be retained.}

\subsection{Partition function and virtual ghosts}
Although there are no negative norm physical states (for the right
range of $d$), the orbifold theory may still contain negative norm
virtual states running in loops. This can be studied by looking
at the one-loop partition function. Before considering the orbifold
case, we recall the partition function of the closed bosonic
string  in $26$-dimensional Minkowski space,\footnote{More precisely, 
the definition of $Z(\tau,\bar{\tau})$ is
\begin{equation}
\nonumber
Z(\tau,\bar{\tau})=\Tr(-1)^F c_0b_0\tilde{c}_0\tilde{b}_0 q^{H_L}\bar{q}^{H_R}
\end{equation}
where $(-1)^F$ anticommutes with all the ghost fields. In the following, 
we implicitly assume that the trace is taken with $(-1)^F c_0b_0\tilde{c}_0\tilde{b}_0$ inserted. 
}
\begin{eqnarray*}
Z(\tau,\bar{\tau})=\Tr q^{H_L}\bar{q}^{H_R} &\sim&
{V_{26}\over\tau_2^{12}}\Bigl({1\over|q^{1\over 24}\prod_m (1-q^m)|^{2}}
\Bigr)^{26} \times |q^{1 \over 24} \prod_m (1-q^m)|^{4}\\
&=& {V_{26}\over\tau_2^{12}}
\Bigl({1 \over|q^{1 \over 24} \prod_m (1-q^m)|^{2}}\Bigr)^{24}\,,
\end{eqnarray*}
where $H_{L}=L_0-a$, $H_{R}=\tilde{L}_0-a$, $a$ is the zero point
energy, and $q=e^{2\pi i \tau}$. $V_{26}$ is a space-time volume
factor related to the continuum normalization of the momentum integral
and $\prod_m$ is a short hand for $\prod_{m=1}^\infty$.  Here, the
contributions from the negative and positive norm ghost states cancel
the contributions from the time-like and one space-like oscillators,
respectively, giving the same result as we would get in the light-cone 
gauge. To verify that this really is how the cancellations work, one
can compute the following closely related quantity,
\begin{eqnarray*}
S(\tau,\bar{\tau})=\Tr(-1)^{s} q^{H_L}\bar{q}^{H_R}
&\sim&
{V_{26}\over\tau_2^{12}}
\frac{|q^{1\over 24}\prod_m (1-q^m)|^{2}\,|q^{1\over 24}\prod_m
(1+q^m)|^{2}}
{\Bigl(|q^{1 \over 24} \prod_m (1-q^m)|^{2}\Bigr)^{25}
\Bigl(|q^{1 \over 24} \prod_m (1+q^m)|^{2}\Bigr)} \\
&=& {V_{26}\over\tau_2^{12}}
\Bigl({1 \over|q^{1 \over 24} \prod_m (1-q^m)|^{2}} \Bigr)^{24}\,.
\end{eqnarray*}
The insertion $(-1)^s$ ensures that negative norm states contribute
with a negative sign in the trace. The equality $Z(\tau,\bar{\tau})=
S(\tau,\bar{\tau})$ then reflects the fact that the negative norm
ghost contribution really did cancel that of the time-like
oscillators.\footnote{We thank C. Vafa for this argument.}

We now proceed to the partition function of the orbifold
$R^{1,d}/Z_2$. A partition function, being a vacuum amplitude, is a
space-time scalar. Therefore, the trace in it extends to the
space-time index of the states, {\it i.e.}, the conjugate to
$\alpha^\mu_{-n}|p\ket$ appears as $\bra p|\alpha_{\mu,n}$. The
commutators then involve $\delta^\mu_\nu$ rather than
$\eta_{\mu\nu}$ and time-like oscillators contribute in the same way
as space-like ones. The partition function then has the same form as
that of the Euclidean orbifold $R^{1+d}/Z_2$, and is given by
\begin{eqnarray}
Z(\tau,\bar{\tau})&=&\Tr{}_U{{1+\hat{g}}\over 2}q^{H_L}\bar{q}^{H_R}
+\Tr{}_T {{1+\hat{g}} \over 2}q^{H_L}\bar{q}^{H_R}\label{partition} \\
&=&
\frac{V_{25-d}}{2}
\left(\frac{1}{\sqrt{\tau_2}\, |q^{1 \over 24} \prod_m (1-q^m)|^{2}}
\right)^{24-(d+1)}     \nonumber \\
&&\times\,\, \Biggl[
\frac{V_{d+1}}{(\sqrt{\tau_2}\,|q^{1\over 24}\prod_m (1-q^m)|^2)^{d+1}}
+\frac{V_{d+1}}{|q^{1 \over 24} \prod_m (1+q^m)|^{2(d+1)}}\nonumber \\
&&\quad
+\frac{1}{|q^{-\frac{1}{48}} \prod_m (1-q^{m-{1\over 2}})|^{2(d+1)}}
+\frac{1}{|q^{-{1\over 48}} \prod_m (1+q^{m-{1\over 2}})|^{2(d+1)}}
\Biggr]\,,
\end{eqnarray}
where $\hat{g}$ is the $Z_2$ action on the Hilbert space and $\Tr_U$ 
and $\Tr_T$ denote traces over untwisted and twisted sector states
respectively.
Note that since twisted sector states do not carry momentum in the
$d+1$ orbifolded directions, their contributions do not contain the
corresponding continuum normalization factor $V_{d+1}$ for the
momentum integral. The partition function is modular invariant with
$V_{d+1}=2^{-(d+1)}$. It is instructive to compare this with the torus
orbifold $T^{d+1}/Z_2$. In that case, since the momenta in the
orbifold directions are discrete, the normalization is trivial. But
then the orbifold has $2^{d+1}$ twisted sectors associated with as
many fixed points which results in modular invariance.

Although the expression for the partition function of the Lorentzian
orbifold $R^{1,d}/Z_2$ is the same as that of the Euclidean orbifold
$R^{d+1}/Z_2$, they embody very different physics. In the Euclidean
orbifold, the negative norm ghost contribution always cancels against
time-like oscillators, indicating the possibility of choosing a gauge
(the light-cone gauge), in which there are no negative norm states.
However, in the Lorentzian orbifold $ R^{1, d}/Z_2$, virtual
negative norm states make uncancelled contributions to the partition function.
As a check
of this, one can again look at
\begin{equation}
S(\tau,\bar{\tau}) =\Tr{}_U(-1)^{s} {{1+\hat{g}} \over
2}q^{H_L}\bar{q}^{H_R} +\Tr{}_T (-1)^{s}{{1+\hat{g}} \over
2}q^{H_L}\bar{q}^{H_R}\,.
\label{sig}
\end{equation}
The difference, if any, between  $S(\tau, \bar{\tau})$ 
and $Z(\tau,\bar{\tau})$ can only arise because of differing 
contributions from the negative norm states and therefore, we
concentrate on these parts of  $Z(\tau,\bar{\tau})$ and $S(\tau,\bar{\tau})$.  
The contribution from negative norm states in the
left moving sector to various terms in (\ref{partition},\ref{sig}) is given
by:
\begin{eqnarray}
\Tr{}_U(\pm1)^{s} q^{L_0}&\sim&
\prod_n{{(1\mp q^n)} \over {(1\mp q^n)}},~~~~~~~
\Tr{}_U(\pm1)^{s} {\hat{g}}q^{L_0}~\sim~
\prod_n{{(1\mp q^n)} \over {(1\pm q^n)}}, \nonumber  \\ 
\Tr{}_T(\pm 1)^{s} q^{L_0}&\sim&
\prod_n{{(1\mp q^n)} \over {(1 \mp q^{n+{1 \over 2}})}},~~~~
\Tr{}_T(\pm 1)^{s} {\hat{g}}q^{L_0}~\sim~
\prod_n{{(1\mp q^n)} \over {(1\pm q^{n+{1 \over 2}})}}. \label{negcontr}
\end{eqnarray}
The upper signs in the four expressions correspond to 
contributions of negative norms states to $Z(\tau,\bar{\tau})$ 
and the lower signs correspond to their contributions
to $S(\tau, \bar{\tau})$. 
The factor $(1\mp q^n)$ in the numerator is the contribution
from the negative norm ghost states and the factor in the denominator
is the contribution from the time-like oscillators.\footnote{Note that
for the Euclidean orbifold, the contributions from the time-like
oscillator is always $\frac{1}{1\mp q^n}$ which cancels the contribution
from the negative norm ghost oscillators resulting in the equality
$Z(\tau,\bar{\tau})=S(\tau,\bar{\tau})$ for the Euclidean orbifold.}
From these expressions, it is clear that the contributions of the
negative norm states to $Z(\tau,\bar{\tau})$ and $S(\tau,\bar{\tau})$
are not the same. Hence
$S(\tau,\bar{\tau})
\neq Z(\tau,\bar{\tau})$,  
which explicitly shows that the contributions of virtual negative norm
states in the partition function do not cancel on the Lorentzian
orbifold. This implies that
 a ghost-free gauge for string theory in such a
background does not exist. This is perhaps not surprising: we cannot choose the light
cone gauge because our orbifold involves a reflection in the time
direction. One might have thought that there is some other gauge in
which all calculations can be done in terms of positive norm states, but
the analysis above shows that such a gauge does not exist. Nevertheless,
as we have shown, there are no negative norm physical states on the 
orbifold. 
We expect that these
features are generic for theories in time dependent
backgrounds.

Since $b$ and $c$ are reparameterization ghosts, their periodicities on
the world-sheet torus are fixed by the theory. However, suppose that we
regard the $X^\mu$ as describing simply a free field theory on the
orbifold. Then, in principle, we can introduce a $(b,c)$ system, not
as reparameterization ghosts, but with the sole purpose of removing the
negative norm states. Then, by assigning appropriate periodicities to
the ghosts, depending on the $X^\mu$ boundary conditions, it is
possible to fully cancel the contributions of  negative norm states 
in all sectors of the theory. However, this will not be a string theory.
\paragraph{Summary}
We have found that there are no negative norm physical states in the
bosonic string theory on the Lorentzian orbifold $R^{1,d}/Z_2$ when
$d+1 \geq 9$, and the partition function is modular invariant. 
However, negative norm virtual states make uncancelled contributions
to quantum loops. This implies that it is not possible to choose
a gauge in which all computations are done in terms of positive
norm states. 
For
$9 \leq d+1 \leq 16$, the ground state in the twisted sector $|p^i
\ket$ carrying momentum in the un-orbifolded directions is physical
with $|\vec{p}|^2={15-d \over 8}$. For $d+1 > 16$, there are no
physical states in the twisted sector.

\section{Type II superstrings on the Lorentzian orbifold}

We will next move on to type II superstrings. Because the orbifold
involves time, we will work in the covariant RNS formulation. Now the
orbifold action is
\be
X^a \rightarrow -X^a\,,\quad X^i\rightarrow X^i\,, \qquad
\psi^a \rightarrow -\psi^a\,, \quad \psi^i \rightarrow \psi^i\,,
\ee
where, $a=0,\cdots, d$ and $i=d+1,\cdots, 9$. For technical 
reason, we will always consider $d$ odd. 

We first look at the
untwisted sector. Here, the fermions have the standard mode expansions:
$\psi^\mu(\sigma_-)=\sum_r\psi^\mu_r~e^{-ir(\tau-\sigma)}$, with
similar expressions for left-movers $\tilde{\psi}^\mu(\sigma_+)$.
The sum is over ${r\in Z+\half}$ in the NS sector and ${r\in Z}$ in
the R sector. The bosons have the mode expansions (\ref{xunt}). The
zero point energy $a=a_B+a_F$ is $a=\half$ in the NS sector and $a=0$
in the R sector. The NS sector ground state is a tachyonic scalar
$|p^a,p^i\ket_{NS}$, whereas the R ground state is a massless spinor
$|p^a,p^i\ket_R$. The orbifold operation acts on the R vacuum as
\be
|p^a,p^i\ket_R\rightarrow (\Gamma_{11})^{d+1}\Gamma^0\Gamma^1
\cdots\Gamma^d\, |-p^a,p^i\ket_R.
\label{Rproj}
\ee
After the orbifold projection, the invariant states have momentum
wave-functions of definite symmetry, $|p^a,p^i\ket\pm|-p^a,p^i\ket$,
depending on the $(d+1)$-dimensional chirality of the R ground state
and the oscillator numbers.

As in the bosonic orbifold of the previous section, the
physical untwisted orbifold states form a subspace in the space of
physical states of the parent type II theory. Consequently, the
untwisted sector is free of physical negative norm states.

The supersymmetry of the physical untwisted spectrum (for odd $d$)
can be illustrated as follows. Let $\CS^{(1,9)}$ denote an $SO(1,9)$
spin-field of definite chirality that relates the NS and R ground
states in type II theory, $|p\ket_{R}\sim\CS^{(1,9)}|p\ket_{NS}$. Then
the space-time supersymmetry current in the unorbifolded theory has
the form $J\sim e^{-\phi/2} \CS^{(1,9)}$, where $e^{-\phi/2}$ is the
spin-field for the $\beta,\gamma$ ghost system. Suppose that
$\CS^{(1,9)}$ has positive chirality and we denote $SO(n)$ spinors
of $\pm 1$ chirality by $\CS^{(n)}_\pm$. As the orbifold breaks
$SO(1,9)$ to $SO(1,d)\times SO(9-d)$, the positive chirality
spin-field decomposes as
$$
\CS^{(1,9)}_+=\CS^{(1,d)}_+\otimes \CS^{(9-d)}_+ +\CS^{(1,d)}_-
\otimes \CS^{(9-d)}_-.
$$
Then, under the orbifold projection (\ref{Rproj}), the piece with
positive $SO(1,d)$ chirality survives and the orbifold inherits a
supercurrent $J_{orb}\sim e^{-\phi/2}\CS^{(1,d)}_+\otimes \CS^{(9-d)}_+$
from the parent theory. This proves the supersymmetry of the untwisted
sector, while showing that the amount of supersymmetry has been
reduced.

We now turn to the twisted sector. The twisted bosons have the
mode expansion (\ref{xtwist}). Fermions in the twisted sector satisfy
boundary conditions:
\bea
{\rm NS}:&& \psi^a(\sigma+2\pi)=\psi^a (\sigma)\,,\qquad
\psi^i (\sigma + 2\pi)=-\psi^i(\sigma)\,; \\
{\rm R}:&&  \psi^a (\sigma+2\pi)=-\psi^a (\sigma)\,,\,\quad
\psi^i (\sigma+2\pi )=\psi^i (\sigma)\,.
\eea
These lead to the mode expansions:
\bea
{\rm NS}:&& \psi^a (\sigma_- )=\sum_{n\in Z}
\psi^a_n~e^{-in(\tau-\sigma)}\,,\quad\! \psi^i(\sigma_-)
=\sum_{r\in Z+\half} \psi^i_r~e^{-ir(\tau-\sigma)}\, ;\\
{\rm R}:&&\psi^a (\sigma_- )=
\sum_{r\in Z+\half} \psi^a_r~e^{-ir(\tau-\sigma)}\,,\quad
\psi^i (\sigma_-)=\sum_{n\in Z} \psi^i_n~e^{-in(\tau-\sigma)}\,.
\eea
The periodicities and mode expansions are reversed along the
orbifolded directions compared to the unorbifolded ones. 
The twisted NS sector has fermion zero modes along the orbifold and
the corresponding ground state $|p^i\ket^T_{NS}$ is a $SO(1,d)$
spinor and a $SO(9-d)$ scalar. The twisted R sector ground state
$|p^i\ket^T_{R}$ is a spinor under $SO(9-d)$ and a scalar under
$SO(1,d)$. Some more details can be found in the Appendix B.

Using the mode expansions, the Virasoro generators $L_m$ and the
worldsheet supercurrents $G_r$ and $F_n$ can be worked out. These are
summarized in Appendix A. To identify the physical spectrum, one also
needs the zero point energies, $a=a_B+a_F$. In the NS sector,
the worldsheet bosonic and fermionic sectors contribute as,
\begin{equation}
a_B=-{d+1 \over 48}+{9-d \over 24}-{2 \over 24}\,,\qquad
a_F=-{d+1 \over 24}+{9-d \over 48}-{2\over 48}\,.
\end{equation}
Here, $-2/24$ is the contribution from the $b,c$ ghosts and
$-2/48$ is the contribution from the NS sector $\beta, \gamma$
ghosts. In the twisted Ramond sector, $a_B$ is as above and the
fermions give,
\begin{equation}
a_F=-{(9-d) \over 24}+{d+1 \over 48}+{2 \over 24}\,,
\end{equation}
where $2/24$ is from the Ramond sector $\beta, \gamma$ ghosts.
In total then,
\bea
&& a=a_B+a_F={3-d \over 8}\qquad (\mbox{Twisted\,, NS})\,,\\
&& a=a_B+a_F=0 \qquad\qquad (\mbox{Twisted\,, R})\,.
\eea
The zero point energy vanishes for any value of $d$ in the twisted
Ramond sector.

\subsection{Twisted sector physical states}

The content of the twisted sector physical spectrum is determined by
the super-Virasoro constraints,
\begin{equation}
\begin{array}{c}
(L_m-a\,\delta_m)|\mbox{phys}\ket=0 \quad (m\geq 0),
\nonumber\\[.3cm]
G_r |\mbox{phys}\ket=0 \quad (r\geq\half\,,\,\,{\rm NS})\,,
\qquad
 F_n |\mbox{phys}\ket=0 \quad (n\geq 0\,,\,\, {\rm R})\,,
\end{array}
\label{conditions}
\end{equation}
with the generators given in Appendix A. As in the bosonic case, the
$L_0$ constraint gives
\bea
p^ip_i + \sum_l l N_l &=& {3-d \over 8}\quad (\mbox{NS sector}),\\
                      &=& 0\quad \qquad (\mbox{R sector})\,.
\eea
Here $p^i$ is the momentum carried by the twisted sector state in the
unorbifolded direction, and $\sum_l l N_l$ schematically represents
the combined sum over the bosonic, fermionic and ghost number
operators in the twisted sector. Note that the minimum
non-zero value of this sum is $\half$, while the right hand side is
always less than $\half$. Therefore, physical twisted states cannot
have any oscillator excitations. In particular, they will be free of
negative norms. In the twisted NS sector, there are no physical states
for $d>3$. For $d \leq 3$, the twisted NS ground state
$|p^i\ket^T_{NS}$ is physical with $p^ip_i={3-d \over 8}$. In
particular, for the case of $d=3$, this ground
state has $p^i=0$. 

This state also trivially satisfies all the other
physical state constraints in (\ref{conditions}). In the R sector, the
only physical state, for any $d$, is the Ramond ground state at zero
momentum, $|p^i=0\ket^T_{R}$. This also satisfies the remaining
constraints in (\ref{conditions}). In particular, the $F_0$ constraint
gives $p_i\Gamma^i|p^i\ket^T_R =0$, which is normally the Dirac
equation reducing the number of spinor components by half. In our
case, since $p^i=0$, it does not impose a constraint. Thus, {\em e.g.} in $d=3$
the twisted R sector vacuum has twice as many components as the
twisted sector NS vacuum (see Appendix B).

The GSO projection results in the NS sector ground state, $|p^i\ket^T_{NS}$, 
having  the same $SO(1,d)$ chirality in the left and right moving 
sectors.\footnote{Recall we consider odd $d$ so chirality
is well defined.}  
In the twisted R sector, the ground state
$|p^i\ket^T_{R}$, has the same (opposite) $SO(9-d)$ chirality in the 
left and right moving sector for Type IIB (Type IIA) string
theory. 

In general, the bosonic and fermionic degrees of freedom in the
twisted sector will not match. For the special case of $d=3$, 
the twisted sector NS ground state is a chiral spinor of $SO(1,3)$ and
the R sector ground state is a chiral spinor of $SO(6)$. These
spinors have different dimensionalities and as a result bose-fermi
degeneracy of the space-time spectrum is broken in the twisted sector.
\footnote{In the case
of the Euclidean orbifold $R^4/Z_2$, the Dirac equation in the
Ramond sector reduced the fermionic components by half resulting
in a supersymmetric spectrum. }

\subsection{Partition function and tadpoles}

The one-loop partition function, as in the bosonic case, does not
distinguish between space-like and time-like oscillators. Therefore,
the result for superstrings on the Lorentzian orbifold $R^{1,d}/Z_2$
will be the same as that for the Euclidean orbifold
$R^{d+1}/Z_2$. This is in spite of the fact that the spectra in the
two cases are very different, especially in the twisted sector. 
 For definiteness, we look
at the case of $d=3$.

The torus partition function for the orbifold is given by
$$
Z(\tau,\bar{\tau})=\Tr {}_{U}{(1+\hat{g}) \over 2}
q^{H_L}\bar{q}^{H_R}+\Tr{}_{T}{(1+\hat{g}) \over 2}
q^{H_L}\bar{q}^{H_R},
$$
where $\hat{g}$ is the representation of the $Z_2$ orbifold action on
the Fock space, and $\Tr_U$ and $\Tr_T$ represent traces taken over
the untwisted and the twisted sectors.  We also need to sum over the
four different spin structures of the torus in both sectors.  The
contributions from the $b,c$ and $\beta,\gamma$ ghosts will cancel the
contributions from two unorbifolded Euclidean directions. Then, for
the $d=3$ case, the result after the relative sign factors for the
contributions from different spin structures have been chosen, is
\begin{eqnarray}
Z\!\!\!\!&=&\!\!\!\!\frac{V_6}{2\tau_2^2\eta^4\bar{\eta}^4}
\sum_{h,g=0}^1 \frac{(V_4)^{1-h} Z_b^{(h,g)}}{(16)^{(h-1)g}}
\times \sum_{a,b=0}^1(-1)^{(a+b+ab)}
{{\theta^2 { \left[\begin{array}{c} a \\ b \end{array}\right]}
\theta \left[ \begin{array}{c} a+h \\ b+g \end{array} \right]
\theta \left[ \begin{array}{c} a-h \\ b-g \end{array} \right] }
\over 2\eta^4} \nonumber \\[.2cm]
&&\qquad\qquad\qquad\qquad
\times \sum_{\bar{a},\bar{b}=0}^1(-1)^{(\bar{a}+\bar{b}+
\lambda\bar{a}
\bar{b})}
{\bar{\theta}^2 \left[ \begin{array}{c} \bar{a}\\ \bar{b} \end{array}
\right] \bar{\theta}\left[ \begin{array}{c} \bar{a}+h \\ \bar{b}+g
\end{array} \right]
\bar{\theta} \left[ \begin{array}{c} \bar{a}-h \\ \bar{b}-g
\end{array} \right]
\over 2\bar{\eta}^4}\,,
\label{Z4}
\end{eqnarray}
where $\lambda = 0,1$ for type IIA, IIB superstring. This is the same
as the Euclidean case (see, for example, \cite{Kiritsis:1997hj}). The
$\theta$-functions are defined as
$$
\theta \left[ \begin{array}{c} a \\ b \end{array} \right] \equiv
\theta \left[ \begin{array}{c} a \\ b \end{array} \right] (0| \tau )
=\sum_{n\in Z}q^{{1\over2}(n-{a\over2})^2} e^{-\pi i b(n-{a\over2})}\,,
$$
and $Z_b$ is the contribution from the bosonic sector,
$$
Z_b^{(0,0)}=\frac{1}{{\tau_2}^2 \eta^4\bar{\eta}^4}\,,\qquad
Z_b^{(h,g)}= \frac{\eta^2\bar{\eta}^2}{\theta^2
\left[ \begin{array}{c} 1-h \\ 1-g \end{array} \right]
\bar{\theta}^2 \left[ \begin{array}{c} 1-h \\ 1-g \end{array}
\right]}\,,\quad (h,g)\neq(0,0).
$$
$V_6$ and $V_4$ are volume factors entering the continuum normalization
of the momentum integrals parallel and transverse to the orbifold. $h=0$
for the twisted sector and $g=0$ for terms without the operator
$\hat g$. The contributions from the worldsheet fermions vanish in
each one of the four $(h,g)$ sectors separately, due to the Jacobi
identity,
$$
\half \sum^1_{a,b=0} (-1)^{a+b+ab} \prod^4_{i=1} \theta \left[
\begin{array}{c}
a+h_i \\ b+g_i \end{array} \right] = -\prod^4_{i=1}
 \theta \left[ \begin{array}{c}
1-h_i \\ 1-g_i \end{array} \right]\,,
\label{Jacobi}
$$
coupled with  $\theta\left[\begin{array}{c} 1\\1\end{array}\right]=0$.
Hence $Z(\tau, \bar{\tau})=0$, without having to fix the relative
factor $V_4$. The vanishing of the partition function in particular
implies its modular invariance.
 All this looks rather surprising
considering the difference between the Euclidean and Lorentzian
orbifolds. As in the bosonic case, the difference can be made manifest
by inserting, in the partition function, an operator $(-1)^s$ that
changes the sign of all negative norm states. Once again one finds
that although the physical spectrum is free of negative norm states,
non-physical negative norm states do not decouple in the loops.

The vanishing of the partition function implies that there is no
dilaton tadpole, at zero energy-momentum \cite{Ginsparg:1986wr}. 
In the absence of
the orbifold, a non-zero momentum tadpole vanishes simply by momentum
conservation. However, in the space-time orbifold, because of
energy-momentum non-conservation at the ``conical'' singularity,
kinematics can allow inserting, on the torus, a dilaton vertex
operator carrying non-zero energy and momentum. The vanishing of such
tadpoles is not obvious and requires further investigation.

\paragraph{Summary}
For Type II superstring, we have found that there are no
negative norm {\em physical} states on the Lorentzian
orbifold $R^{1,d}/Z_2$. The ground state in the twisted
NS sector transforms as a spinor of $SO(1,d)$ and a scalar
of $SO(9-d)$. It is only physical when $d \leq 3$ and the
momentum it carries in the un-orbifolded directions has to
satisfy $|\vec{p}|^2={{3-d}\over 8}$. In the twisted Ramond
sector, the ground state is a $SO(1,d)$ scalar and $SO(9-d)$
spinor. It is physical for any value of $d$ and its momentum
in the un-orbifolded directions has to vanish: $p^i=0$. The
partition function is modular invariant and the zero momentum
dilaton tadpole vanishes at one loop.  

\section{Discussion}
In this article we studied two basic issues in string theory about
which very little is known -- time-dependent backgrounds and
cosmological singularities.  We chose the simplest possible spaces
exhibiting these phenomena, space-time orbifolds of Minkowski space,
and showed how simple quotients by time reversal and spatial
reflections evade some of the obvious potential pitfalls (tachyons and
ghosts in the physical spectrum, 
zero-momentum tadpoles at one loop, lack of modular
invariance etc.). Although there are closed time-like loops
in the construction, quantum mechanical evolution is consistent
because the orbifold prescription projects onto states that are 
invariant under the discrete identification.

How is an S-matrix defined when a class of physical states is
localized in time?  An asymptotic observer in models such as ours only
observes transition amplitudes between the propagating untwisted
sector states.  Any such amplitude could involve the emission of
arbitrarily many twisted sector states which cannot be observed at
late or early times.  Therefore it appears that the rules for
computing transition amplitudes in space-time orbifolds will require
tracing over emissions of states that are localized in time.  If so,
pure states scattering off a space-time orbifold singularity could
emerge as mixed states due to entanglement with an unobservable
twisted sector.  Perhaps such a mechanism is responsible for creating
the observed large entropy of the universe in a cosmological context.

One might wonder whether a loss of unitarity is also implied by the
uncancelled contributions of negative norm virtual states in the
partition function of our space-time orbifolds.  Certainly, this result
implies that it is not possible to choose a ghost-free gauge in which
all computations are carried out in terms of positive-norm states.  We
expect that this will be true in many time-dependent backgrounds of
string theory -- it is at least clear that the ghost-free light-cone
gauge cannot be chosen in time-dependent backgrounds.  This is in
sharp contrast to usual string theories and field theories in static
backgrounds.  Nevertheless, it is not clear that a loss of unitarity
in transition amplitudes is implied.  In particular, since our models
do not have any {\it physical} negative norm states, cutting the one
loop diagram will not give a transition amplitude to a ghostlike
state.  In the absence of a general argument connecting negative norm
virtual contributions to the partition function and S-matrix
unitarity, we require detailed study of amplitudes for propagating
untwisted sector states scattering from the orbifold singularity. 

There are very interesting subtleties in the computation of
correlation functions and transition amplitudes on space-time
orbifolds such as ours in which the twisted sectors are localized in
time.  Because of the localization, we do not expect energy (or
momentum in any of the orbifolded directions) to be conserved in
interactions between the untwisted and twisted sectors.  One important
consequence is that (unlike usual spatial orbifolds) kinematics does
not forbid a finite momentum tadpole appearing at one loop.  We can 
expect that this issue of finite-momentum tadpoles will persist for 
time-dependent string backgrounds in general.

One reason for our focus on the $R^{1,3}/Z_{2}$ orbifold of the
superstring is that this orbifold had ``massless'' twisted sector
states with Euclidean momenta satisfying $\vec{p}^{2} = 0$. In the
classic $R^{4}/Z_{2}$ orbifold the massless twisted sector states (for
which Lorentzian $\vec{p}^{2} = 0$) correspond to geometric blowup
modes which can resolve the singularity. Some condensates of the
twisted sector states correspond to parameters of the Eguchi-Hanson
Ricci-flat metric on the smooth manifold obtained by replacing the tip
of the $R^{4}/Z_{2}$ cone by a sphere. We might hope that some conical
space-time singularities can be resolved by similar condensates of
``massless'' twisted sector states. When the twisted sector states are
tachyonic, we might similarly hope that tachyon condensation would
resolve the orbifold singularity. Unfortunately, much of the geometric
technology of deforming singular manifolds into smooth spaces relies
on complex geometry and cannot accommodate a manifold with signature
$(1,d)$. For example, the Eguchi-Hanson metric \cite{Eguchi} has
signature $(4,0)$ and while one can easily obtain a $(2,2)$ signature
Ricci-flat metric by analytic continuation from it, a (1,3) signature
Ricci-flat metric cannot be obtained in this way.\footnote{A simple
generalization of the Eguchi-Hanson metric cannot work because the
curvature two form for an Eguchi-Hanson space is self dual, but in
$(1,3)$ signature, the self-duality condition has an extra factor of
'$i$'.} In order to understand cosmological singularities in string
theory it is urgent that we develop the mathematics of resolution of
singularities of Lorentzian manifolds.\footnote{Perhaps the geometric
difficulty in resolving these singularities is related
to the fact that the ``massless'' twisted sector states
are not exactly moduli fields in the low energy theory. 
They are localized in time and are on shell only at zero
momentum. }

In string theory, the quantum mechanics of a relativistic string is
used to compute transition amplitudes and an S-matrix for the
scattering of conventional multi-graviton states.  In view of this we
have studied the quantum mechanics of strings on space-time orbifolds. 
Field theories on such spaces raise several new issues.  For example,
new singularities can potentially arise in correlation functions of
operators at space-like separations if the space-time interval between
some operators and the orbifold images of others is time-like or null.  
The rules for defining field theories in such backgrounds remain to be 
worked out.\footnote{We are grateful to Nati Seiberg for a discussion 
of these issues.  Also see~\cite{natietal}.}

We conclude here by summarizing some perspectives from this work
about time-dependent backgrounds and cosmological singularities in
string theory:
\begin{itemize}
    \item String theories defined  on time dependent backgrounds run 
    the risk of having ghosts and tachyons in the physical spectrum.    
    \item Even when there are no ghosts in the physical spectrum, 
    negative norm states can make uncancelled contributions to the 
    partition function.  In such cases it is not possible to choose a 
    ghost-free gauge like light-cone gauge.  This might lead to loss of 
    unitarity, but a more detailed analysis is needed.    
    \item The quantum mechanics of strings on space-time orbifolds can 
    be consistently defined even if there are closed time-like loops by
    projecting onto states invariant under the orbifold group. It would be
interesting to consider space-time orbifolds without closed time-like
curves, but we expect the issues raised here to persist (see \cite{natietal}).
    \item The resulting orbifolds can be tachyon and ghost-free and 
    typically contain a twisted sector at a fixed plane localized in 
    time.
    \item Scattering from such an asymptotically unobservable twisted 
    sector could cause transitions from a pure state to a mixed 
    state, generating entropy.
    \item Since energy need not be conserved in a time-dependent 
    background, kinematics does not forbid the production of tadpoles 
    with finite momentum.   Hence, the vanishing of these amplitudes
    must be checked to confirm the existence of a valid solution to 
    string theory.
\end{itemize}
We expect to return to many of the issues laid out above in a future
publication. 

\bigskip

\noindent
{\bf \large Acknowledgments}

\bigskip

We thank J. Christian, M. Cvetic, E. Gimon,  P. Kraus, D. Minic, B. Ovrut, N.
Seiberg, G. Shiu, C. Thorn, C. Vafa, and E. Witten for useful
conversations. {\small E.K-V.} has been in part supported by the
Academy of Finland, and thanks the University of Pennsylvania for
hospitality. {\small V.B.} dedicates this work to 
Ravi Nicholas who arrived while our paper was being written.
 {\small V.B.} and {\small A.N.} were supported by DOE
grant DOE-FG02-95ER40893. 

\appendix
\section{Twisted sector superconformal generators}
Here we list the super conformal generators in the twisted sector of
the $R^{1,d}/Z_2$ orbifold theory. The Virasoro generators are given
by
\be
L^{NS}_m = L^{B}_m + L^{F,NS}_m\,,\qquad
L^{R}_m  = L^{B}_m + L^{F,R}_m
\ee
where, in the twisted sector,
\bea
L^{B}_m
&=& \half \sum_{n\in Z} : \left( \alpha^a_{-n-\half}
\alpha^b_{m+n+\half} \eta_{ab} + \alpha^i_{-n}\alpha^i_{m+n}\right) :
\eea
and
\bea
L^{F,NS}_m &=& \half \sum_{n\in Z} (n+\frac{m}{2})
: \psi^a_{-n}\psi^b_{m+n} \eta_{ab} : +\half \sum_{r\in Z+\half}
(r+\frac{m}{2}) :\psi^i_{-r}\psi^i_{m+r} :
\\
L^{F,R}_m &=& \half \sum_{r\in Z+\half} (r+\frac{m}{2}) : \psi^a_{-r}
\psi^b_{m+r} \eta_{ab} : + \half \sum_{n\in Z} (n+\frac{m}{2})
: \psi^i_{-n}\psi^i_{m+n} :
\eea
The super current components in the twisted sector are
\bea
 G_r &=& \sum_m :\left( \psi^a_{-m}\alpha^b_{r+m}\eta_{ab} +
 \psi^i_{r+m}\alpha^i_{-m} \right) : \\
 F_n &=& \sum_m :\left( \psi^a_{-m+\half}
 \alpha^b_{n+m-\half}\eta_{ab} + \psi^i_{n+m}\alpha^i_{-m}
 \right) : \ .
\eea
There are similar expressions for the left-movers.
\section{Twisted sector vacua as spinors}
Consider $2n$ worldsheet fermion zero modes $\psi^a_0$ satisfying
$\{\psi^a_0,\psi^b_0\}=\delta^{ab}$ and commuting with the mass
operator. The theory then has an $SO(2n)$ spinor as its degenerate
vacuum, which can be constructed as follows. Define
$\Gamma^a=\sqrt{2}\psi^a_0$. These then satisfy the Dirac algebra
$\{\Gamma^a ,\Gamma^b\}=2\delta^{ab}$. The ground state is a
representation of this algebra and can be constructed using the
standard procedure: For $k=1,\cdots n$, define
$$
e_k=\half(\Gamma_k + i \Gamma_{n+k})\,,\qquad
e^\dagger_k=\half(\Gamma_k - i \Gamma_{n+k})\,.
$$
These satisfy the fermionic algebra, $\{e^\dagger_k,e_l\}=
\delta_{kl}$, with other anti-commutators vanishing. Start from a
state $|0\ket$ annihilated by the lowering operators. Other components
of the ground state spinor are obtained by using the raising operators
on this lowest state; $|0\ket, e^\dagger_k|0\ket, e^\dagger_ke^
\dagger_l|0\ket, \cdots, e^\dagger_1e^\dagger_2\cdots^\dagger_n|0\ket$.
The degeneracy of a state with $p$ raising operators is the
combinatoric factor, $^nC_p$ and the total number of states is $2^n$;
that of a spinor. The chirality operator is
$\Gamma_{2n+1}=\Gamma_1\cdots\Gamma_{2n}$. States with even (odd) number
of oscillators form a positive (negative) chirality spinor. In the
twisted NS sector, $2n=d+1$ and in the twisted R sector $2n=9-d$.

\end{document}